# Graph analysis and modularity of brain functional connectivity networks: searching for the optimal threshold

Cécile Bordier, Carlo Nicolini, Angelo Bifone

**Abstract**

Neuroimaging data can be represented as networks of nodes and edges that capture the topological organization of the brain connectivity. Graph theory provides a general and powerful framework to study these networks and their structure at various scales. By way of example, community detection methods have been widely applied to investigate the modular structure of many natural networks, including brain functional connectivity networks. Sparsification procedures are often applied to remove the weakest edges, which are the most affected by experimental noise, and to reduce the density of the graph, thus making it theoretically and computationally more tractable. However, weak links may also contain significant structural information, and procedures to identify the optimal tradeoff are the subject of active research. Here, we explore the use of percolation analysis, a method grounded in statistical physics, to identify the optimal sparsification threshold for community detection in brain connectivity networks.

By using synthetic networks endowed with a ground-truth modular structure and realistic topological features typical of human brain functional connectivity networks, we show that percolation analysis can be applied to identify the optimal sparsification threshold that maximizes information on the networks' community structure. We validate this approach using three different community detection methods widely applied to the analysis of brain connectivity networks: Newman's modularity, InfoMap and Asymptotical Surprise. Importantly, we test the effects of noise and data variability, which are critical factors to determine the optimal threshold. This data-driven method should prove particularly useful in the analysis of the community structure of brain networks in populations characterized by different connectivity strengths, such as patients and controls.

**Introduction**

In recent years, considerable efforts have been made to study the complex structure of brain connectivity, marking the inception of the "connectomic era" in brain neuroscience. Functional Magnetic Resonance Imaging (fMRI) and other neuroimaging methods have shown that spontaneous fluctuation in brain activity, as measured with a subject lying in the scanner without being engaged in any specific task, are organized in coherent patterns, thus suggesting that resting state functional connectivity reflects the functional architecture of the brain (1).

Several methods have been developed and applied to study these patterns of synchronization, including multivariate approaches (e.g. Principal Component or Independent Component Analysis) (1, 2) and graph theoretical methods (3).

Graph theory provides a general and powerful framework to investigate the topological organization of the brain connectivity. A number of graph theoretical studies have revealed a small-world, rich-club structure (4) of functional connectivity networks, and the presence of hub regions defined by high connectivity and network centrality. Moreover, community detection methods have been widely applied to investigate the modular structure of many natural networks, including brain functional connectivity networks. The presence modules, i.e. clusters of nodes that are more densely connected among them than with the rest of the network, reflects functional segregation within the integrated network, and is thought to confer robustness and adaptability to brain connectivity networks (3).

For these analyses, the brain is represented as a network of nodes interconnected by links. Commonly, the nodes correspond to anatomically defined brain areas and links to a measure of inter-regional interaction or similarity between the nodes. For resting state functional connectivity networks, edge weights are typically computed as temporal correlations in the fluctuations of the BOLD signals in different areas, resulting in a correlation adjacency matrix (5).

Sparsification procedures are normally applied to remove weaker links, which are most affected by experimental noise (6), and to reduce the density of the graph, thus making it computationally more tractable. In the literature, it is common practice to fix the density of the adjacency matrix *a priori*, and to identify the threshold that preserves the target density of edges (7, 8). Stability analyses exploring a range of densities are often performed to assess how critically topological parameters derived from the sparsified adjacency matrix depend on the choice of threshold. Sparsification schemes based on the computation of graph Minimum Spanning Trees prior to thresholding have also been proposed to prevent disruption of local connectivity by global removal of weak links (9) .

Here, we address the problem of computing the optimal threshold for community detection in brain connectivity networks. Specifically, we propose the use of percolation analysis, a method rooted in statistical physics, to identify a sparsification threshold that maximizes information on the network modular structure. This data driven procedure, first introduced by Gallos et al. (10), iteratively removes the weakest edges and computes the largest connected component. The percolation threshold corresponds to the point where the largest component starts breaking apart. We entertain the hypothesis that the percolation threshold strikes the optimal balance between information gained by cutting off noise, and lost by removing potentially genuine weak connections. To test this hypothesis, we apply three different community detection methods (Newman's modularity (11), InfoMap (12, 13) and Asymptotical Surprise (14, 15)) to synthetic networks endowed with a ground truth modular structure, and with topological features, levels of noise and variability similar to those observed in functional connectivity experimental data. We compare the retrieved and planted modular structures by using Normalized Mutual Information, an information theoretic measure of similarity, as a function of sparsification threshold. We find that this information can be maximized by an appropriate choice of threshold, and we assess the use of percolation analysis as a data-driven method for optimal sparsification. Finally, we discuss the application of this approach to compare networks characterized by different noise levels and connectivity strengths, such as those observed in cross-sectional studies assessing brain connectivity in different populations, e.g. patients and healthy controls.

**Materials and Methods**

Synthetic networks are a useful tool to test the effect of threshold on community detections, and the ability to retrieve a pre-determined ground-truth modular structure. We ran two types of simulations: simulation of Lancichinetti-Fortunato-Radicchi (LFR) networks (16)  and simulation of complex LFR including intersubject variability and different level of noise. The latter made it possible to assess the influence of noise or data variability, thus mimicking realistic experimental dataset.
The main goal of these simulations is the validation of a method that can be used in the analysis of functional connectivity networks as measured by resting state fMRI. As shown in (14), brain functional connectivity networks are composed of modules with heterogeneous size distributions. This structure can be mimicked using the LFR approach, which can generate synthetic networks with power law degree distributions and community sizes akin to those observed in natural networks, such as functional connectivity networks (16).

*Simulation 1*

The Lancichinetti-Fortunato-Radicchi (LFR) benchmark algorithm generates networks with *a priori* known communities and node degree distributions. Community size and node degree follow power law distributions (for example see Fig.1).

The mixing of the communities is controlled by the topological mixing parameter $\mu_t$. Each node shares a fraction $1-\mu_t$ of edges with nodes in its same community and a fraction $\mu_t$ with nodes in other communities: $0 \leq \mu_t \leq 1$. Similarly, a weight mixing coefficient $\mu_w$ controls, on average for each node, the balance between the incident edge weights coming from internal and external communities.

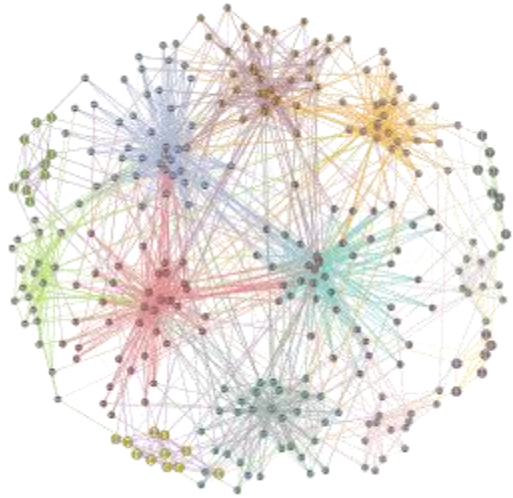

Figure 1: Example of benchmark LFR network with parameters N=300, ⟨k⟩ = 12, $\max_k$ = 50, $\mu_t$ = $\mu_w$=0.2, $\min_c$=5, $\max_c$=50.

The LFR synthetic networks were built for N = 600 nodes, sampling nodes degree from a power-law with exponent $\tau_d$ = 2, average degree ⟨k⟩ = 12 and maximum degree $\max_k$ = 50. We set the topological and weight mixing coefficient, i.e. the average fraction of intra-cluster and intercluster degree and strengths, to $\mu_t = \mu_w = 0.2$. Planted community sizes ranged from 5 to 50 nodes and were sampled from a power law with exponent $\tau_c$ = 1. This simulation was run 9 times for each value of $\mu_t = \mu_w$. The Matlab code to generate LFR synthetic network is available at https://github.com/carlonicolini/lfrwmx. The function takes the parameters described above as inputs, and returns the equivalent to a weighted connectivity matrix that can be directly analyzed by community detection approaches.

*Simulation 2*

This simulation makes use of the output matrix from the LFR function described above to generate artificial resting state fMRI datasets. The general idea is that, starting from an adjacency matrix with a given modular structure, we can generate time-courses for each of the nodes whose pairwise correlations reproduce the edge structure of the original matrix. Schematic of this procedure is shown in Fig.2.

To this end, we first calculate the closest positive-definite matrix to the original LFR network (17), and then we exploit the properties of the Cholesky decomposition and the techniques described in (15) to calculate time-course for the individual nodes. This approach makes it possible to generate correlated random variables, i.e. following the weights/correlations of our original connectivity matrix $C \in R^{n \times n}$, by decomposing it such that $C = LL^T$ with $L \in R^{n \times n}$ and multiplying to it random time series $X \in R^{n \times 1}$ such as $Y = LX$ such that $E(YY^T) = C$ to obtain the desired correlation between the simulated time series. The random time series were generated with 150 points and a base-line value set to 100.

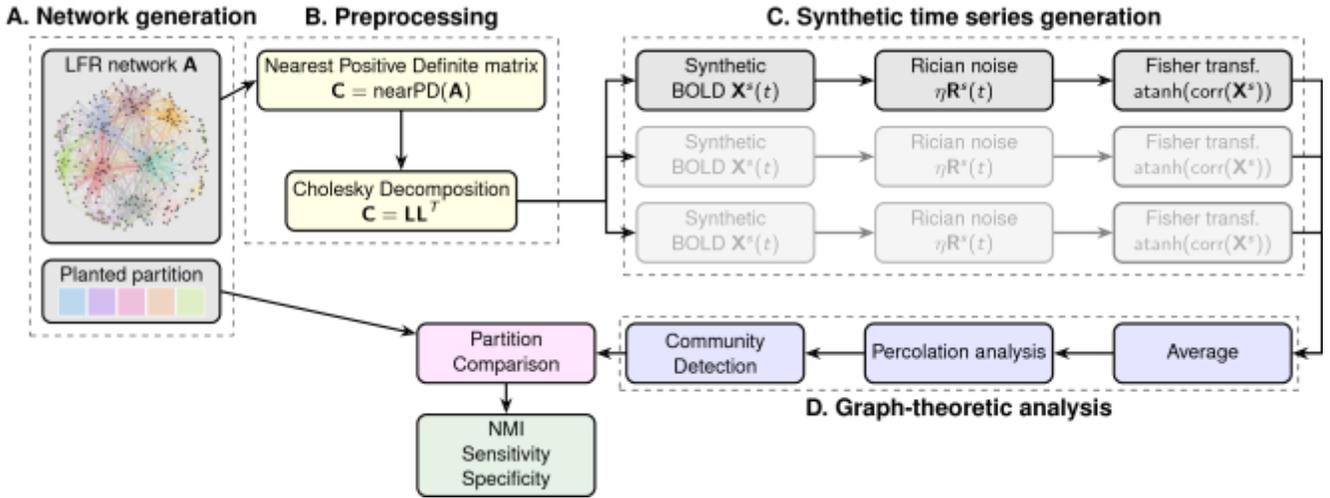

**Figure 2:** Flowchart of the generation and analysis of the synthetic datasets. In A a network with a pre-defined community structure is generated. The adjacency matrix is then processed in block B to obtain the nearest positive definite matrix for the Cholesky decomposition. This enables the generation of node-wise time-courses into which different levels of noise can be injected. The procedure is repeated multiple times to generate different instances (mimicking different subjects in the sample). Finally, correlation matrices are calculated for each instance (block C), and Fisher transformed to calculate the average adjacency matrix for analysis by community detection algorithms (block D). The resulting partitions are then compared with the original, planted one in terms of NMI.

The generation of multiple sample of random time series simulates the effects of intersubject variability, and Rician-noise (18) is added to mimic fMRI resting state data. The definition of Signal-to-Noise (SNR) used in the rest of this paper is: SNR= $\bar{S}/\sigma_N$ where $\bar{S}$ is the average magnitude of the signal and $\sigma_N$ is the standard deviation of the noise (19). Both the random resting state time series and the added noise were generated using the R package NeuRosim (20).

The last step results in 600 times series of 150 points with different levels of noise for each of the simulated subjects. Datasets were generated for populations of 20, 40 and 60 subjects and for a SNR equal to 35 and 70. The procedure has been run 5 times for each different parameter to produce different networks and datasets.

*Connectivity matrix.*

The connectivity matrix is the weighted matrix representing the links between two nodes. In Simulation 1, the matrix was generated directly by the LFR model. In Simulation 2, using the same approach as in fMRI experiment, we computed pairwise Pearson correlations between time-series from pairs of nodes in each dataset (subject), resulting in a matrix M of size N x N with N the number of nodes and with M(i,j) the correlation coefficient between the time series of the node i and the node j. Average group matrices were calculated by Fisher transformation and subsequent averaging of individual matrices.

*Sparsification and Percolation Threshold*

Sparsification procedures are normally applied to remove weaker links, which are most affected by experimental noise (6), and to reduce the density of the graph, thus making it computationally more tractable.

The method of our choice for the sparsification was motivated by a model to describe phase transitions of connected subgraphs in random networks called percolation analysis (21, 22). We applied thresholds on the original network at different levels of edge weights, and identified the largest connected components of the thresholded graphs via breadth-first search (23). The critical point where the largest component starts breaking apart is identified as the percolation threshold at which the network's structure, is preserved while discarding potential effects of noise. Fig.3 represents an example of the size of the largest component with respect to the threshold in a benchmark LFR network.

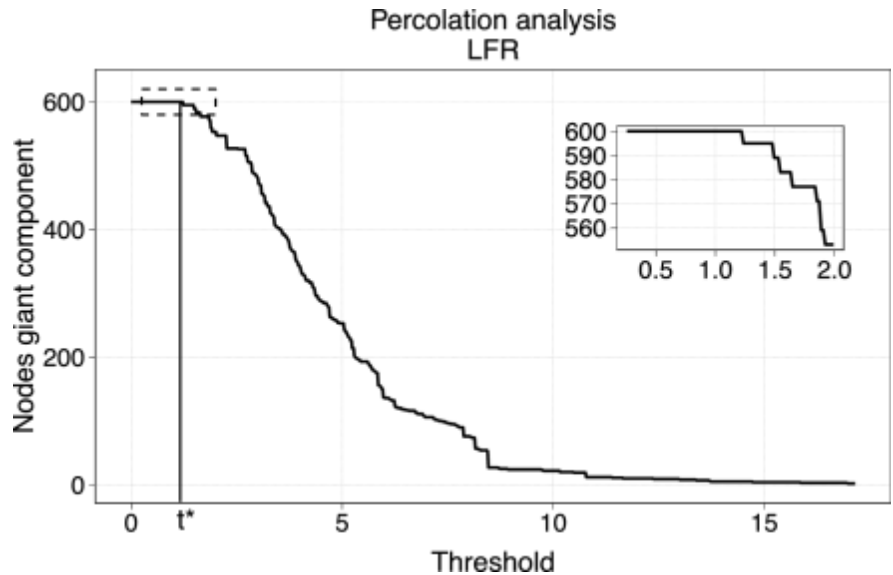

Figure 3: Percolation analysis for a LFR networks. The number of nodes in the giant component has a step-wise behaviour with respect to the threshold. The percolation threshold value is t*.

*Community detection*

To assess whether the efficacy of the sparsification procedure depends on the community detection approach, we applied three different methods, based on conceptually different principles that have been extensively applied to the analysis of resting state fMRI data. The first one, probably the most widely used, is Newman's modularity (11). We also tested InfoMap (12) and Asymptotical Surprise (14, 15), as they have been shown to resolve community structures at a finer level than Newman's modularity, which is affected by a resolution limit that prevents detection of modules that are smaller than a scale determined by the size of the entire network.

Briefly, Newman's modularity seeks optimal partition by maximizing intra-cluster edge-density against that of a null model based on random edge rewiring. Optimization of this fitness function is typically performed using the Louvain method (24), a greedy agglomerative clustering algorithm that works on hierarchical refinements of the network's partitions. Here we used the Louvain implementation available in the Brain Connectivity toolbox (25)).

The idea behind Infomap is the minimization, through a set of heuristics, of the description length (26) of a random walker defined on the network. For this study we used the Infomap implementation available in the igraph-0.7.1 package (27).

Finally, Asymptotical Surprise is a recently developed approach rooted in information theory that aims at maximizing the relative entropy between the observed intracluster density and the expected intracluster density, on the basis of the Erdos-Renyi null model (28). Surprise was recently shown to be quasi-resolution-limit free, and to provide improved means to resolve the modular structure of complex networks of brain functional connectivity (14, 15). Optimization of Asymptotical Surprise was carried out by means of PACO (PArtitioning Cost Optimization), an iterative agglomerative algorithm built on a variation of the Kruskal algorithm for minimum spanning trees(14, 15). We have shown that maximization of Asymptotical Surprise enables detection of communities of widely different sizes, thus making it possible to resolve differences in the modular organization of different networks representing functional connectivity in different subjects or experimental groups(14). A Matlab toolbox including binary and weighted versions of Surprise optimization is available upon request at http://forms.iit.it/view.php?id=68447.

*Evaluation of retrieved partition*

The advantage to know in advance the ground truth community is that we can quantify differences between the planted community and the extracted ones. Three coefficients were used to evaluate the results of the community detection methods at different levels of threshold of our synthetic networks. First, the Normalized Mutual Information (NMI)(29, 30), a measure of the similarity between structures is defined as:

$$NMI(A,B) = \frac{-2 \sum_{i=1}^{C_A} \sum_{j=1}^{C_B} N_{ij} \log\left(\frac{N_{ij} N}{N_{i.} N_{.j}}\right)}{\sum_{i=1}^{C_A} N_{i.} \log\left(\frac{N_{i.}}{N}\right) + \sum_{j=1}^{C_B} N_{.j} \log\left(\frac{N_{.j}}{N}\right)}$$

where $A$ and $B$ are the community structures of two networks, $C_A$ and $C_B$ are the number of community in partition $A$ and $B$ respectively, $N$ the total number of nodes in the networks (which is the same in $A$ and $B$) and $N_{ij}$ is the overlap between $A$'s community $i$ and $B$'s community $j$; i.e. the number of common nodes. Finally, $N_{i.}$ and $N_{.j}$ are the total number of nodes in community $i$ of $A$ and $j$ of $B$ respectively. The NMI ranges from 0 to 1, where 0 indicates that the retrieved community structure does not convey information about the planted partition, and 1 when the two partitions correspond perfectly.

In order to gain information about the origin of mismatches between planted and retrieved partitions, we also computed Sensitivity and Specificity, assessing the levels of false positives and false negatives incurred by the community detection algorithms. For each community we identified the biggest overlap between the ground truth and the retrieved modules to establish a correspondence between the partitions. Subsequently, we identified the nodes that were correctly assigned (true positives=TP) and wrongly assigned (false positives=FP) to a selected community. We also identified nodes that were correctly assigned (true negatives=TN) or erroneously assigned (false negative= FN) to a different community. These values were used to calculate Sensitivity and Specificity for each community:

$$Sensitivity = \frac{TP}{TP + FN}$$

$$Specificity = \frac{TN}{TN + FP}$$

and subsequently averaged over the partition. The values for Sensitivity and Specificity range from 0 to 1, with 1 denoting the perfect match.

**Results**

*Simulation 1*

The benchmark created for this first test did not involve any variation coming from noise or subject variability. The community detections methods were applied directly to the matrix generated by the LFR function. Figure 4 shows the NMI calculated between the structure extracted by Newman modularity, InfoMap and Asymptotical Surprise, and the ground truth for $\mu_t = \mu_w = 0.2$, as a function of threshold.

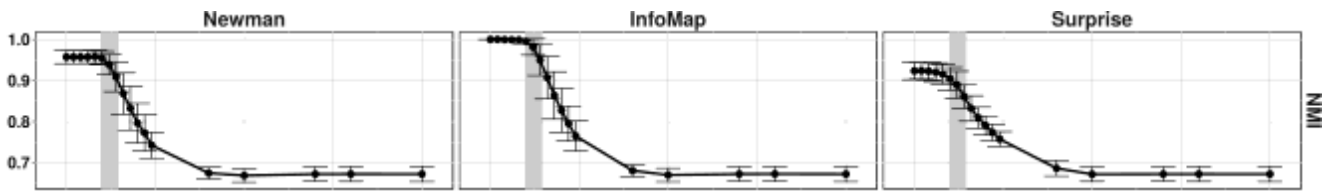

**Figure 4: NMI between ground truth community structure and the results of the 3 community detection algorithms applied to an LFR networks ($\mu_t=\mu_w=0.2$)**

The gray zone on the graphics indicates the range of sparsification thresholds obtained by percolation analysis calculated in different runs. These graphics demonstrate the deleterious effects of excessive removal of weak edges. In the case of noiseless networks, percolation analysis identifies the threshold corresponding to the departure from optimal performance of the community detection algorithm. This is in keeping with the fact that the percolation threshold is the minimum threshold value that preserves connectedness of the giant component. However, it should be noticed in this noiseless scenario all links correspond to true correlations, and no spurious edges are contemplated.

*Simulation 2*

In the second simulation we assessed the effects of noise and variability in the correlation structure of the networks. We computed NMI, the Sensitivity and Specificity for the partitions obtained by the 3 methods (Newman, InfoMap and Asymptotical Surprise) with the different SNRs and numbers of subjects (see Fig. 5, 6 and7).

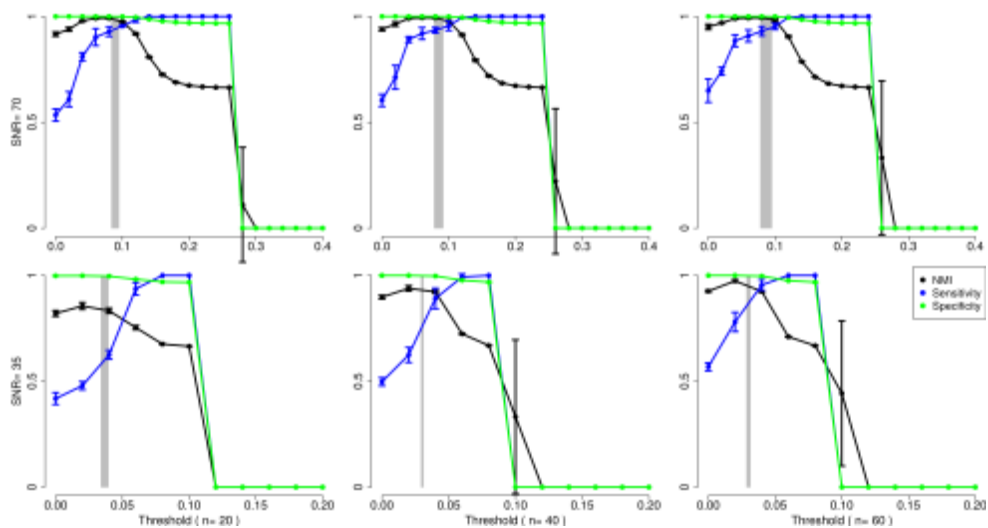

**Figure 5: NMI (in black), Sensitivity (in blue) and Specificity (in green) of the Newman community detection algorithm applied to LFR networks ($\mu_t=\mu_w=0.2$). Two different signal to noise ratio (SNR) are represented on the lines (top line SNR=70, lower line SNR=35), Number of subjects varies depending of the column (respectively from left to right 20, 40, 60 subjects)**

In the presence of variability, we observe a first increase in NMI for increasing threshold, followed by a subsequent drop. We interpret the first rise as a regime in which weak links are mostly determined by spurious correlations, and carry little information about the structure of the network. As threshold increases, removal of additional edges decreases the ability to retrieve the planted modular structure by removing structurally relevant correlations. This picture is confirmed by the observation that maxima in NMI correspond to simultaneously large values of Sensitivity and Specificity.

The percolation threshold values appear to consistently fall in the vicinity of maximum NMI for all three community detection methods.

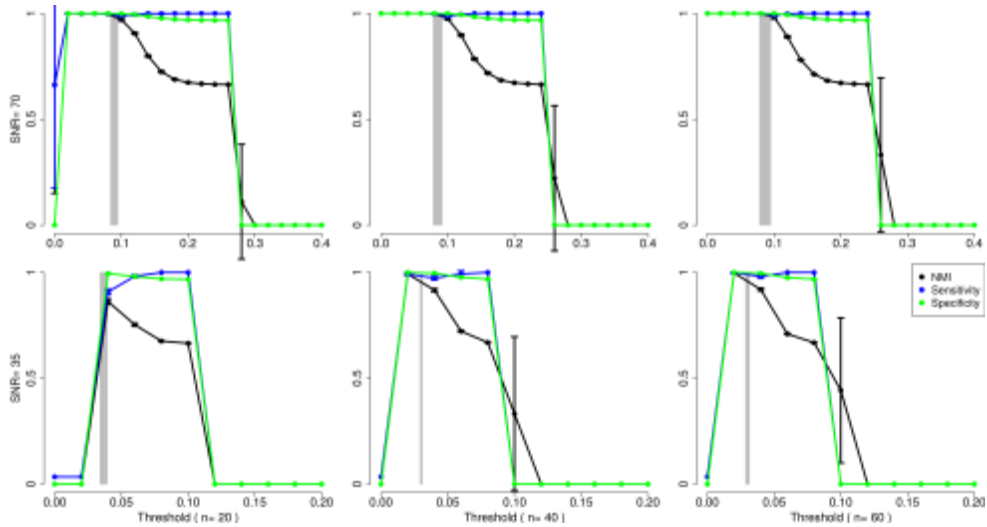

**Figure 6:** NMI (in black), Sensitivity (in blue) and Specificity (in green) of the InfoMap community detection algorithm applied to LFR networks ($\mu_t=\mu_w=0.2$). Two different signal to noise ratio (SNR) are represented on the lines (top line SNR=70, lower line SNR=35), Number of subjects varies depending of the column (respectively from left to right 20, 40, 60 subjects)

The general conclusion from these figures is that percolation analysis is able to detect a quasi-optimal value of sparsification threshold, thus enabling optimal detection of community structure in the presence of experimental noise and data variability.

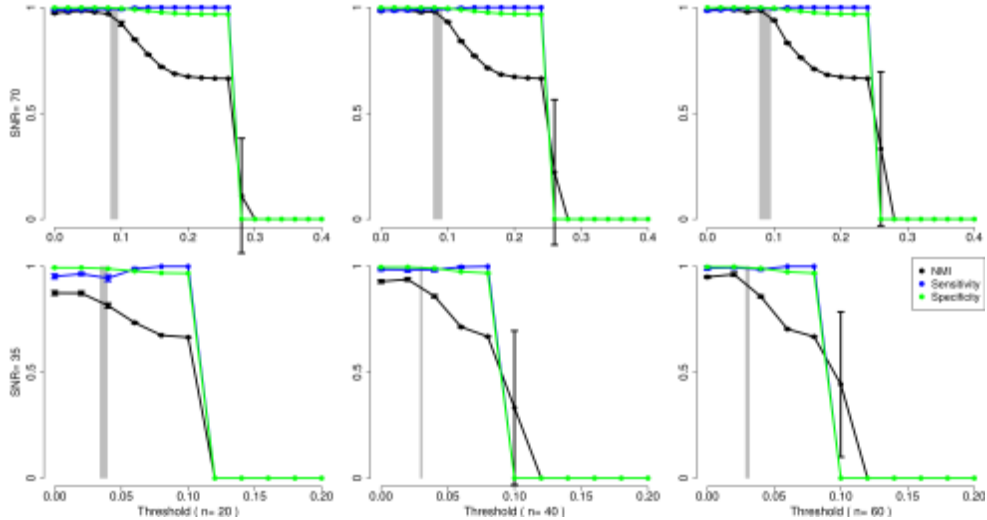

**Figure 7:** NMI (in black), Sensitivity (in blue) and Specificity (in green) of the Asymptotical Surprise community detection algorithm applied to LFR networks ($\mu_t=\mu_w=0.2$). Two different signal to noise ratio (SNR) are represented on the lines (top line SNR=70, lower line SNR=35), Number of subjects varies depending of the column (respectively from left to right 20, 40, 60 subjects)

## Discussion

An open problem in the analysis of brain connectivity is the optimal choice of threshold when comparing different groups, e.g. patients and healthy controls in cross-sectional studies assessing the effects of disease on functional connectivity. Typically, identical sets of nodes are defined for the two groups, and the comparison is based on edge distribution and strength. Many studies tend to fix the same edge density in the connectivity graphs of the groups to be compared. Indeed, certain global topological parameters (e.g. global efficiency (25)) depend on edge density, and comparisons at constant density make it possible to assess differences related to the topological reorganization of links, rather than to their number and strength. On the other hand, constant edge density may bias group comparisons when graphs exhibit intrinsic differences in connectivity strength. By way of example, neuropsychiatric diseases like Schizophrenia and Autism have been associated with disruption and overall reduction of functional and structural connectivity. Imposing equal densities for graphs describing connectivity in patients and controls may lead to the inclusion of a greater number of weak, potentially spurious links in the group with weaker connectivity, and to the exclusion of important links in the group with stronger connectivity. A higher proportion of spurious connection results in a more random network topology, and intergroup differences may just reflect different levels of noise, rather than genuine topological differences (6).

The present study may provide a strategy to overcome this problem. Indeed, community detection determines the membership of each node to a certain module. This is not dependent on overall edge density, but on the local balance between edges linking the node to other members of the same module, or to other nodes in different modules. The optimal sparsification threshold is the one that maximizes information about community structure, and is network-specific, as it depends on the structure and noisiness of each network. Hence, independent thresholding of the networks to be compared based on percolation analysis maximizes information about memberships in the two groups.

**Conclusion**

In conclusion, we have explored the use of percolation analysis, a method based on statistical physics, to determine the sparsification threshold in synthetic networks endowed with a ground-truth modular structure, and topological features akin to those of real world networks like brain connectivity graphs. We find that the percolation threshold, i.e. the highest threshold that preserves connectedness of the giant component, corresponds to the maximum information that can be retrieved by various community detection algorithms on the planted modular structure in the presence of noise and intersubject variability. Intuitively, this threshold corresponds to the optimal balance between information lost by removing genuine edges and spurious correlations introduced by noise. These findings provide evidence of the existence of an optimal sparsification threshold, and a solid theoretical basis for its identification by means of a data driven method like percolation analysis.

**Acknowledgements**

This project has received funding from the European Union's Horizon 2020 Research and Innovation Program under grant agreement No 668863

**References**

1. Damoiseaux JS, et al. (2006) Consistent resting-state networks across healthy subjects. *Proc Natl Acad Sci U S A* 103(2):13848–53.

2. Beckmann CF, DeLuca M, Devlin JT, Smith SM (2005) Investigations into resting-state connectivity using independent component analysis. *Philos Trans R Soc London - Ser B Biol Sci* 360(1457):1001–1013.

3. Bullmore ET, Sporns O (2009) Complex brain networks: graph theoretical analysis of structural and functional systems. *Nat Rev Neurosci* 10(3):186–198.

4. van den Heuvel MP, Sporns O (2011) Rich-Club Organization of the Human Connectome. *J Neurosci* 31(44):15775–15786.

5. Eguiluz VM, Chialvo DR, Cecchi GA, Baliki M, Apkarian A V (2005) Scale-free brain functional networks. *Phys Rev Lett* 94(1):18102.

6. van den Heuvel M, Fornito A (2014) Brain Networks in Schizophrenia. *Neuropsychol Rev* 24(1):32–48.

7. Bassett DS, et al. (2008) Hierarchical organization of human cortical networks in health and schizophrenia. *J Neurosci* 28(37):9239–48.

8. Lynall ME, et al. (2010) Functional connectivity and brain networks in schizophrenia. *J Neurosci* 30(28):9477–9487.

9. Alexander-Bloch AF, et al. (2010) Disrupted Modularity and Local Connectivity of Brain Functional Networks in Childhood-Onset Schizophrenia. *Front Syst Neurosci* 4.

10. Gallos LK, Makse HA, Sigman M (2012) A small world of weak ties provides optimal global integration of self-similar modules in functional brain networks. *Proc Natl Acad Sci U S A* 109(8):2825–2830.

11. Newman ME (2006) Modularity and community structure in networks. *Proc Natl Acad Sci U S A* 103(23):8577–8582.

12. Rosvall M, Bergstrom CT (2008) Maps of random walks on complex networks reveal community structure. *Proc Natl Acad Sci U S A* 105(4):1118–23.

13. Kawamoto T, Rosvall M (2015) Estimating the resolution limit of the map equation in community detection. *Phys Rev E* 91(1):10.

14. Nicolini C, Bifone A (2016) Modular structure of brain functional networks: breaking the resolution limit by Surprise. *Sci Rep* 6(19250).

15. Nicolini C, Bordier C, Bifone A (2017) Modular organization of weighted brain networks beyond the resolution limit. *Neuroimage* 146:28–39.

16. Lancichinetti A, Fortunato S, Radicchi F (2008) Benchmark graphs for testing community detection algorithms. *Phys Rev E* 78(4):46110.

17. Higham NJ (1988) Computing a nearest symmetric positive semidefinite matrix. *Linear Algebra Appl* 103(C):103–118.

18. Welvaert M, Rosseel Y (2013) On the definition of signal-to-noise ratio and contrast-to-noise ratio for fmri data. *PLoS One* 8(11):e77089.

19. Krüger G, Glover GH (2001) Physiological noise in oxygenation-sensitive magnetic resonance imaging. *Magn Reson Med* 46(4):631–637.

20. Welvaert M, Durnez J, Moerkerke B, Verdoolaege G, Rosseel Y (2011) neuRosim: An R package for generating fMRI data. *J Stat Softw* 44(10):1–18.

21. Callaway DS, Newman MEJ, Strogatz SH, Watts DJ (2000) Network robustness and fragility:

percolation on random graphs. *Phys Rev Lett* 85(25):5468–5471.

22. Goerdt A (2001) The giant component threshold for random regular graphs with edge faults. *Theor Comput Sci* 259(1–2):307–321.

23. Leiserson CCE, Rivest RRL, Stein C, Cormen TH (2009) *Introduction to Algorithms, Third Edition* doi:10.2307/2583667.

24. Blondel VD, Guillaume J-L, Lambiotte R, Lefebvre E (2008) Fast unfolding of communities in large networks. *J Stat Mech Theory Exp* 10008(10):6.

25. Rubinov M, Sporns O (2010) Complex network measures of brain connectivity: Uses and interpretations. *Neuroimage* 52(3):1059–1069.

26. Rissanen J (1978) Modeling by shortest data description. *Automatica* 14(5):465–471.

27. Csárdi G, Nepusz T (2006) The igraph software package for complex network research. *InterJournal Complex Syst* 1695:1695.

28. Traag VA, Aldecoa R, Delvenne JC (2015) Detecting communities using asymptotical surprise. *Phys Rev E - Stat Nonlinear, Soft Matter Phys* 92(2). doi:10.1103/PhysRevE.92.022816.

29. Danon L, Guilera AD, Duch J, Arenas A (2005) Comparing community structure identification. *J Stat Mech Theory Exp* 2005(9):P09008--9008.

30. Meilă M (2007) Comparing clusterings-an information based distance. *J Multivar Anal* 98(5):873–895.